\documentclass[aps,pre,preprint,showpacs,eqsecnum,amssymb]{revtex4}
\usepackage{graphicx,color,psfrag}
\begin{document}
\date{\today}
\title{Large deviations in the alternating mass harmonic chain}
\author{Hans C. Fogedby}
\email{fogedby@phys.au.dk}
\affiliation{Department of Physics and
Astronomy, University of
Aarhus, Ny Munkegade\\
8000 Aarhus C, Denmark\\}
\affiliation{Niels Bohr Institute, Blegdamsvej 17\\
2100 Copenhagen {\O}, Denmark}
\begin{abstract}
We extend the work of Kannan et al. and derive the cumulant generating function for
the alternating mass harmonic chain consisting of $N$ particles and driven by heat reservoirs.
The main result is a closed expression for the cumulant generating function in the thermodynamic
large $N$ limit. This expression is independent of $N$ but depends on whether the chain consists 
of an even or an odd number of particles, in accordance with the results obtained by Kannan el al.
for the heat current. This result is in accordance with the absence of local thermodynamic equilibrium
in a linear system.
\end{abstract}
\pacs{05.40.-a, 05.70.Ln}.

\maketitle
\section{\label{intro} Introduction}
There is a current interest in fluctuating small systems in contact
with heat reservoirs and driven by external forces. This focus is driven
by the recent possibilities of direct manipulation of nano systems 
and bio molecules. These techniques also permit direct experimental
access to the probability distributions for the work or heat exchanged 
with the environment  \cite{Trepagnier04,Collin05,Seifert06a,Seifert06b,Wang02,
Imparato07,Ciliberto06,Ciliberto07, Ciliberto08}. Moreover, these single molecule techniques
have also yielded access to the so called fluctuation theorems,
which relate the probability of observing entropy-generated trajectories, 
with that of observing entropy-consuming trajectories
\cite{Jarzynski97,Kurchan98,Gallavotti96,Crooks99,Crooks00,
Seifert05a,Seifert05b,Evans93,Evans94,Gallavotti95,
Lebowitz99,Gaspard04,Imparato06,
vanZon03,vanZon04,vanZon03a,vanZon04a,Seifert05c}.
As a result there is a general renewed theoretical interest in small
non equilibrium systems.

In the context of non equilibrium systems the well-known 
fluctuation-dissipation theorem, relating response to fluctuations close
to equilibrium has been generalized to the so-called asymptotic 
fluctuation theorem (AFT) valid also far from equilibrium
\cite{Evans93,Evans94,Gallavotti95,Lebowitz99,
Kurchan98,Crooks99,Seifert05a}. The AFT,
which has been demonstrated under quite general conditions, implies 
for the cumulant generating function  (CGF) the fundamental 
symmetry
\begin{eqnarray}
\mu(\lambda)=\mu(\beta_1-\beta_2-\lambda). 
\label{aft}
\end{eqnarray}
The CGF, $\mu(\lambda)$,  is defined at long times $t$ according to
\begin{eqnarray}
\langle\exp(\lambda Q(t)\rangle\sim\exp(t\mu(\lambda)),
\label{charheat}
\end{eqnarray}
where $Q(t)$ is the accumulated heat transferred to the system
from a reservoir in the time span $t$. Here $\beta_1=1/T_1$ and $\beta_2=1/T_2$ are the
inverse temperatures of the heat reservoirs driving the non equilibrium
process and $\langle\cdots\rangle$ denotes a non equilibrium ensemble average. 
Normalization implies $\mu(0)=0$ and the AFT in (\ref{aft}) in
particular yields $\mu(\beta_1-\beta_2)=0$. In general $\mu(\lambda)$
is a downward convex function passing through $\lambda=0$ and
$\lambda=\beta_1-\beta_2$. $\mu(\lambda)$ is, moreover, bounded by
branch points at $\lambda_+$ and $\lambda_-$. 

Recently, there has been focus on the explicit evaluation of
$\mu(\lambda)$ for deterministic systems driven by Langevin type
heat bath in order to verify the AFT and at the same time determine
how system dependent properties enter in the form of
$\mu(\lambda)$. Little is known about $\mu(\lambda)$ in the case
of interacting or random systems and the focus has therefore been
on tractable linear systems. In a series of papers Saito and Dhar
and Kundu et al. \cite{Saito07,Kundu11}, see also
\cite{Sabhapandit11,Sabhapandit12,Fogedby12}, have discussed the
driven harmonic chain. Here one finds that $\mu(\lambda)$ is a
functional of
\begin{eqnarray}
f(\lambda)=T_1T_2\lambda(\beta_1-\beta_2-\lambda),
\label{f}
\end{eqnarray}
where inspection reveals that $f(\lambda)$ is invariant under the
AFT symmetry in (\ref{aft}). The functional in the case of the
deterministic harmonic linear chain has the generic form
\begin{eqnarray}
\mu(\lambda)=-\frac{1}{2}\int\frac{d\omega}{2\pi}\ln\bigg[1+T(\omega)f(\lambda)\bigg],
\label{cgf1}
\end{eqnarray}
where $T(\omega)$ is a model dependent transmission matrix. In 
linear systems the heat is transported ballistically. Local
equilibrium cannot be established and Fourier's law does not hold
 \cite{Bonetto00}. This is reflected in the form of $\mu(\lambda)$ which
is independent of the system size.

In the case of a simple $N$ particle unit mass harmonic chain with inter particle
coupling $\kappa$, attached to walls at the ends, and driven by two heat
reservoirs with common damping $\Gamma$, one obtains the
transmission matrix \cite{Saito07,Kundu11,Fogedby12}
\begin{eqnarray}
&&T(\omega)=(2\Gamma\omega)^2|G_{1N}(\omega)|^2,
\label{tn}
\\
&&G_{1N}(\omega)=\frac{\kappa\sin p}
{\Omega^2\sin(N-1)p-2\kappa\Omega\sin(N-2)p+\kappa^2\sin(N-3)p},
\label{gn}
\\
&&\Omega=-\omega^2+2\kappa-i\Gamma\omega,
\label{Om}
\\
&&\omega^2=4\kappa\sin^2p/2.
\label{disp}
\end{eqnarray}
Here (\ref{tn}) defines $T$ in terms of the the transmission end-to-end
Green's function $G_{1N}$ given in (\ref{gn}). 
The above results have been analyzed in detail in
\cite{Saito07,Kundu11,Fogedby12}. Here we just
remark that the ballistic lattice waves transporting the heat give rise to
the resonance structure in the denominator in (\ref{gn}). The coupling to the
heat reservoirs only enters in $\Omega$ in (\ref{Om}). Finally,
the wave number $p$ is confined to the first Brillouin zone
$|p|\leq\pi$, yielding the frequency band $|\omega|\leq
2\sqrt\kappa$.

A natural and simple extension of the equal mass harmonic chain is the
harmonic chain with alternating masses. In condensed matter
this is the well-known case of a phonon system with a basis. In
this case the dispersion law (\ref{disp}) breaks up into an acoustic
branch and an optical branch, see e.g. \cite{Ashcroft76}. In both
case the heat is transmitted ballistically but shared between the
acoustic and optical phonons. 

In recent work Kannan et al. \cite{Kannan11} have considered this case
and  have in detail analysed the non equilibrium steady state of an alternating mass harmonic 
chain \cite{Casher71,Oconnor74}; further references to work
on the alternating mass chain can be found in \cite{Kannan11}. 
Like in the equal mass case \cite{Rieder67,Nakazawa70},
the  position and momentum steady state
distribution exhibits a Gaussian form with correlation matrix given by
the static position-position, position-momentum, and
momentum-momentum correlations. Defining the local kinetic 
temperature $T_n$ according to (note that $k_{\text{B}}=1$)
$T_n=\langle p_n^2\rangle/m_n$, where $p_n$ is the momentum 
and $m_n$ the mass associated with the n-th site, Kannan et al. find, surprisingly, 
that the local temperature $T_n$ oscillates with period 2; these oscillations even persist
in the thermodynamic limit; in the equilibrium case for $T_1=T_2=T$ the local
temperature $T_n$ locks onto $T$ in accordance with the equipartion theorem. 
Kannan et al. \cite{Kannan11} also discuss the thermodynamic limit
$N\rightarrow\infty$ and find exact expressions for the local temperature profile $T_n$
and the heat current $\langle Q\rangle/t$. They also find that these expressions depend 
on whether the chain is composed of an even or odd number of particles. 

In the present paper we extend the work of Kannan et al. regarding the 
large $N$ limit of the heat current and discuss the cumulant generating 
function $\mu(\lambda)$ (CGF). The CGF yields the full heat distribution in the
long time limit; note that the heat current is given by
the first term in a cumulant expansion of (\ref{charheat}), i.e., $\langle Q\rangle/t=
(d\mu(\lambda)/d\lambda)_{\lambda=0}$. Referring to the results in
\cite{Fogedby12,Saito07,Kundu11},  the central quantity in the evaluation of the 
CGF is the transmission Green's function $G_{1N}(\omega)$  describing the 
propagation of ballistic modes across a chain of size $N$. We consider the CGF and determine 
the transmission matrix $T(\omega)$ entering in the expression (\ref{cgf1}).
As anticipated the structure of $T(\omega)$ exhibits
the two branch structure of the phonon spectrum, both the acoustic and optical branches
contributing to $T(\omega)$. Finally, we derive a close expression for the CGF
in the large $N$ limit. This constitutes the main and new result in the present paper.
In accordance with the absence of local thermodynamic
equilibrium the large $N$ expression for the CGF is manifestly independent of $N$.

The paper is organised in the following manner. In Sec.~\ref{mod}
we present the model under scrutiny, i.e., the alternating mass chain. In Sec.~\ref{anal}
we set up the necessary analytical apparatus. In Sec.~\ref{green}
we introduce the transmission end-to-end Greens function which incorporates the model
dependent component of the CGF. 
Section ~\ref{cum} contains the main result in the present paper, namely a derivation
of the CGF in the large $N$ limit. Section ~\ref{dis} is devoted to a
discussion of the CGF in the large $N$ limit. For completion this section also includes
a discussion of the transmission matrix  and the large deviation function.  Section 
~\ref{con} is devoted to a conclusion. The issue of deriving an expression for the 
end-to-end Green's function suitable for our needs is deferred to an Appendix.
\section{\label{mod} Model}
We consider an alternating mass harmonic chain attached to a wall at the end points.
The spring constant is denoted by $\kappa$ and the two masses in the unit cell are
$m$, the smaller mass, and $M$, the larger mass. The end particles are driven
by heat reservoirs at temperatures $T_1$ and $T_2$, characterised by the
damping constant $\Gamma$. Kannan et al.  \cite{Kannan11} use a determinantal
approach in analysing the stochastic dynamics. We have found it convenient
for our purposes to use an equation of motion approach.We readily distinguish 
two cases depending on the 
boundary conditions. In case A an integer number of unit cells fits in between the walls,
corresponding to an even number of particles; in case B a half unit cell is in contact with
the right wall, corresponding to an odd number of particles. In Fig.~\ref{fig1} we have 
depicted the two cases and the appropriate unit cell.

Denoting the displacement of the particle with mass $m$ in the n-th unit cell by $u_n$ and the displacement
of the particle with mass $M$ by $w_n$, we obtain in bulk the coupled equations of motion
\begin{eqnarray}
&&m\ddot u_n=\kappa(w_n+w_{n-1}-2u_n),
\label{eq1}
\\
&&M\ddot w_n=\kappa(u_n+u_{n+1}-2w_n);
\label{eq2}
\end{eqnarray}
here a dot denotes a time derivative.

Case A (even): From Fig.~\ref{fig1}  it follows that the coupling to the heat reservoirs is described
by the Langevin equations
\begin{eqnarray}
&&m\ddot u_1=-\Gamma\dot u_1+\kappa(w_1-2u_1)+\xi_1,
\label{lan1}
\\
&&M\ddot w_N=-\Gamma\dot w_N+\kappa(u_N-2w_N)+\xi_N,
\label{lan2}
\end{eqnarray}
where $N$ is the number of unit cells, corresponding to $2N$ particles of either mass,
i.e., an even number of particles.

Case B (odd): According to Fig.~\ref{fig1}, (\ref{lan2}) is replaced by
\begin{eqnarray}
M\ddot u_{N+1}=-\Gamma\dot u_{N+1}+\kappa(w_N-2u_{N+1})+\xi_N,
\label{lan3}
\end{eqnarray}
for $N$ unit cells with  only one particle of mass $m$ in the
$(N+1)$-th unit cell, corresponding to an odd number of particles. Finally, the 
noises $\xi_1$ and $\xi_N$, characterising the heat reservoirs, are correlated according to
\begin{eqnarray}
&&\langle\xi_1(t)\xi_1(t')\rangle=2\Gamma T_1\delta(t-t'),
\label{n1}
\\
&&\langle\xi_N(t)\xi_N(t')\rangle=2\Gamma T_2\delta(t-t'). 
\label{n2}
\end{eqnarray}
The above equations of motion define the dynamics of the chain
and the stochastic coupling to the heat reservoirs at temperatures
$T_1$ and $T_2$.

Focussing on the reservoir at temperature $T_1$ the
fluctuating force is given by $-\Gamma\dot u_1+\xi_1$. Consequently, the rate of work 
or heat flux has the form, denoting $Q\equiv Q_1$,
\begin{eqnarray}
\dot Q=\dot u_1(-\Gamma\dot u_1+\xi_1).
\label{flux}
\end{eqnarray}
With respect to the CGF the central quantity in the analysis is,
however, the total heat transmitted to the system during a finite time interval
$t$, i.e.,
\begin{eqnarray}
Q(t)=\int_0^td\tau\dot u_1(-\Gamma\dot u_1+\xi_1).
\label{totheat}
\end{eqnarray}
The heat $Q(t)$ is fluctuating and the issue is to determine its
probability distribution $P(Q,t)=\langle\delta(Q-Q(t))\rangle$,
where $\langle\cdots\rangle$ denotes an ensemble average with
respect to $\xi_1$ and $\xi_N$. In terms of the characteristic
function $\langle\exp(\lambda Q(t)\rangle$ we have by a Laplace
transform \cite{Gradshteyn65}
\begin{eqnarray}
P(Q,t)=\int_{-i\infty}^{i\infty}\frac{d\lambda}{2\pi i}e^{-\lambda
Q}\langle e^{\lambda Q(t)}\rangle,
\label{pdis}
\end{eqnarray}
where at long times $\langle\exp(\lambda Q(t)\rangle\sim\exp(t\mu(\lambda))$.
Note that $Q(t)$ is unbounded and only the time scaled heat
$Q(t)/t$ is endowed with large deviation properties
\cite{Touchette09,Hollander00}. 
\section{\label{anal} Analysis}
The heat reservoirs drive the chain into a stationary non
equilibrium state. The heat is transported ballistically
by the acoustic and optical phonons. The only damping 
mechanism is associated with the heat reservoirs and
sets a time scale given by $1/\Gamma$. Consequently, 
at long times compared to $1/\Gamma$ we can ignore the
initial preparation and employ the Fourier transform,
\begin{eqnarray}
&&u_n(t)=\int\frac{d\omega}{2\pi}\exp(-i\omega t)u_n(\omega),
\label{u}
\\
&&w_n(t)=\int\frac{d\omega}{2\pi}\exp(-i\omega t)w_n(\omega).
\label{w}
\end{eqnarray}
Proceeding with an equation of motion approach, introducing
\begin{eqnarray}
&&\tilde\Omega_1=-m\omega^2+2\kappa, 
\label{tom1}
\\
&&\tilde\Omega_2=-M\omega^2+2\kappa,
\label{tom2}
\end{eqnarray}
the bulk equation of motion (\ref{eq1}) and (\ref{eq2}) take the form
\begin{eqnarray}
&&\tilde\Omega_1u_n=\kappa(w_n+w_{n-1}), 
\label{eq11}
\\
&&\tilde\Omega_2w_n=\kappa(u_n+u_{n+1}).
\label{eq22}
\end{eqnarray}
Commonly, for systems with periodic boundary conditions one searches
for plane wave solutions of the form $u_n, w_n\sim\exp(ipn)$ and readily
finds the dispersion laws for acoustic and optical phonons \cite{Ashcroft76}. In the present 
context for a finite chain coupled to heat baths, it is more convenient to
proceed in a renormalisation group fashion by diluting the degrees of freedom.
Thus eliminating every second site we obtain from (\ref{eq11}) and (\ref{eq22})
bulk equations referring to each separate sublattice,
\begin{eqnarray}
&&\tilde\Omega_1\tilde\Omega_2 u_n=\kappa^2(u_{n+1}+u_{n-1}+2u_n),
\label{eq111}
\\
&&\tilde\Omega_1\tilde\Omega_2 w_n=\kappa^2(w_{n+1}+w_{n-1}+2w_n).
\label{eq222}
\end{eqnarray}
Searching for plane wave solutions, $u_n, w_n\sim\exp(ipn)$,  we find the common dispersion law for the
two sublattices,
\begin{eqnarray}
\tilde\Omega_1\tilde\Omega_2=2\kappa^2(1+\cos p),
\label{disp2}
\end{eqnarray}
or inserting $\tilde\Omega_1$ and $\tilde\Omega_2$ from (\ref{tom1}) and  (\ref{tom2}) the two branches
\begin{eqnarray}
&&\omega_1^2=\kappa\frac{m+M-s}{mM},
\label{ac}
\\
&&\omega_2^2=\kappa\frac{m+M+s}{mM},
\label{op}
\\
&&s=\sqrt{m^2+M^2+2mM\cos p}.
\label{ss}
\end{eqnarray}
For the acoustic branch the $\omega$ range is $0<|\omega|<\sqrt{2\kappa/M}$; 
for the optical branch $\sqrt{2\kappa/m}<|\omega|<\sqrt{2\kappa(m+M)/mM}$.
The wave number range is $|p|\leq\pi$. The dispersion laws, moreover, locks
$\omega$ onto $p$ in the Fourier integrals over $\omega$ in for example
(\ref{u}) and (\ref{w}).

For later purposes we also need the inverse density of phonon
states
\begin{eqnarray}
&&\rho_1(p)=\frac{d\omega_1}{dp}=
\frac{\kappa\sin p}{2\omega_1 s},
\label{invdens1}
\\
&&\rho_2(p)=\frac{d\omega_2}{dp}=
\frac{\kappa\sin p}{2\omega_2 s},
\label{invdens2}
\end{eqnarray}
referring to the acoustic and optical branches, respectively. In Figs.~\ref{fig2} and \ref{fig3} 
we have depicted the phonon
dispersion laws and the inverse density of states as function 
of $\omega$, showing the gap between 
the acoustic and optical branches. We have chosen
the parameter values $m=1$, $M=2$, and $\kappa=1$.

Introducing
\begin{eqnarray}
&&\Omega_1=-m\omega^2+2\kappa-i\omega\Gamma, 
\label{om1}
\\
&&\Omega_2=-M\omega^2+2\kappa-i\omega\Gamma.
\label{om2}
\end{eqnarray}
the coupling to the reservoirs in case A (even) and B (odd) is described by
\begin{eqnarray}
&&\Omega_1u_1=\kappa w_1 +\xi_1,
\label{c1a}
\\
&&\Omega_2w_N=\kappa u_N +\xi_N,
\label{c2a}
\end{eqnarray}
and
\begin{eqnarray}
&&\Omega_1u_1=\kappa w_1 +\xi_1,
\label{c1b}
\\
&&\Omega_1u_{N+1}=\kappa w_N +\xi_N,
\label{c2b}
\end{eqnarray}
respectively.

The clamping of the chain to the walls gives rise to a dynamical coupling of the
$u$ and $w$ sublattices. The $u$ and $w$ displacements driven by the 
noise inputs at the ends can be expressed in the form
\begin{eqnarray}
&&u_n^{\text{A},\text{B}}=G_{n1}^{\text{A},\text{B}}\xi_1+G_{nN}^{\text{A},\text{B}}\xi_N,
\label{un}
\\
&&w_n^{\text{A},\text{B}}=F_{n1}^{\text{A},\text{B}}\xi_1+F_{nN}^{\text{A},\text{B}}\xi_N,
\label{wn}
\end{eqnarray}
where $G^{\text{A},\text{B}}$ and $F^{\text{A},\text{B}}$ are Green's functions describing the propagation
of ballistic modes from the end points to the n-th unit cell in the two cases. 
\section{\label{green} Transmission Green's function}
Since according to (\ref{flux}) we inject heat at the site $u_1$ the relevant transmission end-to-end Green's
function is $G_{1N}^{\text{A},\text{B}}$. In the appendix we have, using an equation of motion
approach, evaluated the Green's functions entering in (\ref{un}) and (\ref{wn}). 
We stress that our derivation is in complete agreement with Kannan et al. \cite{Kannan11},
who use an equivalent determintal approach. Extracting the expressions from the appendix 
with a slight change in notation ($D^{\text{A}}=D_1^{\text{A}}$) we have
\begin{eqnarray}
&&G_{1N}^{\text{A}}(\omega)=\frac{\kappa_{\text{A}}(\omega)\sin p}{D^{\text{A}}(\omega)},
\label{ga}
\\
&&G_{1N}^{\text{B}}(\omega)=\frac{\kappa_{\text{A}}(\omega)\sin p}{D^{\text{B}}(\omega)}.
\label{gb}
\end{eqnarray}
We note that the end-to-end Green's functions have the same structure as (\ref{gn}) in the case
of a simple chain. From the appendix  we also have
\begin{eqnarray}
&&D^{\text{A}}(\omega)=\Omega_{\text{A}}\Omega_{\text{C}}\sin(N-1)p-(\Omega_{\text{A}}\kappa_{\text{D}}+\Omega_{\text{C}}\kappa_{\text{A}})\sin(N-2)p+\kappa_{\text{A}}\kappa_{\text{D}}\sin(N-3)p,
\label{da}
\\
&&D^{\text{B}}(\omega)=\Omega_{\text{A}}^2\sin(N-1)p-2\Omega_{\text{A}}\kappa_{\text{A}}\sin(N-2)p+\kappa_{\text{A}}^2\sin(N-3)p,
\label{db}
\end{eqnarray}
where
\begin{eqnarray}
&&\Omega_{\text{A}}=\Omega_1-\kappa^2/\tilde\Omega_2,
\label{pp1}
\\
&&\Omega_{\text{C}}=(\Omega_2/\kappa)(\tilde\Omega_1-\kappa^2/\tilde\Omega_2)-\kappa,
\label{pp2}
\\
&&\kappa_{\text{A}}(\omega)=\kappa^2/\tilde\Omega_2,
\label{pp3}
\\
&&\kappa_{\text{D}}(\omega)=\kappa\Omega_2/\tilde\Omega_2,
\label{pp4}
\end{eqnarray}
Here the frequency dependent parameters $\tilde\Omega_1$,  $\tilde\Omega_2$, $\Omega_1$, and $\Omega_2$
are given by (\ref{tom1}), (\ref{tom2}), (\ref{om1}), and (\ref{om2}).

We note that the end-to-end Green's functions have a complex $\omega$ dependence owing
to the coupling of the two sublattices, yielding a frequency dependent coupling strength
$\kappa_{\text{A}}(\omega)$ as well as denominators $D^{\text{A}}(\omega)$ and 
$D^{\text{B}}(\omega)$ depending on whether we are in case A (even) or case B (odd). 
Detailed inspection reveals that these expressions are identical to corresponding 
expressions in Kannan et al. \cite{Kannan11}. A tedious analysis, setting $m=M$ 
and noting that the number of particles is twice the number of unit cells, also leads 
to the expression (\ref{gn}) to (\ref{disp}) for the simple chain.
\section{\label{cum} Cumulant generating function in the large $N$ limit}
The derivation of the cumulant generating function (CGF) proceeds like
in \cite{Fogedby12}, see also \cite{Saito07,Kundu11}, with some added technicalities due
to the two band structure. The central model-dependent quantity is the end-to-end Green's functions
given in (\ref{ga}) and (\ref{gb}). Here $\kappa_{\text{A}}(\omega)$ in (\ref{pp3}) is an effective frequency 
dependent coupling strength; the denominators $D^{\text{A}}(\omega)$ and
$D^{\text{B}}(\omega)$ in (\ref{da}) and (\ref{db}) describe the resonance structure of the chain.

The cumulant generating function CGF is given by the generic expression (\ref{cgf1}). In the
present case, referring to case A (even) and case B (odd), we have 
\begin{eqnarray}
&&\mu^{\text{A},\text{B}}(\lambda)=-\frac{1}{2}\int\frac{d\omega}{2\pi}\ln\bigg[1+T^{\text{A},\text{B}}(\omega)f(\lambda)\bigg],
\label{cgf2}
\\
&&f(\lambda)=T_1T_2\lambda(\beta_1-\beta_2-\lambda),
\label{f1}
\end{eqnarray}
where the transmission matrix has the form
\begin{eqnarray}
T^{\text{A},\text{B}}(\omega)=(2\Gamma\omega)^2|G^{\text{A},\text{B}}_{1N}(\omega)|^2;
\label{tom}
\end{eqnarray}
note that in (\ref{cgf2}) the $\omega$ integration is over both acoustic and optical bands.
Inserting  the density of states (\ref{invdens1}) and (\ref{invdens2})
we obtain in particular
\begin{eqnarray}
\mu^{\text{A},\text{B}}(\lambda)=-\int_0^{\pi}\frac{dp}{2\pi}\sum_{n=1,2} 
\rho_n\ln\bigg[1+(2\Gamma\omega_n)^2|G^{\text{A},\text{B}}_{1N}(\omega_n)|^2\bigg],
\label{cgf3}
\end{eqnarray}
where $n=1,2$ refers to the acoustic and optical bands, respectively.

In order to extract the $N$ dependence of the CGF we note that $N$ only
enters in the denominators $D^{\text{A},\text{B}}$ in (\ref{da}) and (\ref{db}).
We proceed using the method in \cite{Fogedby12,Kannan11,Roy08a} by expressing 
$D^{\text{A},\text{B}}$ in the compact form 
\begin{eqnarray}
D^{\text{A},\text{B}}=\tilde A^{\text{A},\text{B}}\sin(N-1)p+\tilde B^{\text{A},\text{B}}\sin(N-2)p+\tilde C^{\text{A},\text{B}}\sin(N-3)p,
\label{d1}
\end{eqnarray}
where from (\ref{da}) and (\ref{db})
\begin{eqnarray}
&&\tilde A^{\text{A}}=\Omega_{\text{A}}\Omega_{\text{C}},~~\tilde B^{\text{A}}=-\Omega_{\text{A}}\kappa_{\text{D}}-\Omega_{\text{C}}\kappa_{\text{A}},~~\tilde C^{\text{A}}=\kappa_{\text{A}}\kappa_{\text{D}},
\label{parA}
\\
&&\tilde A^{\text{B}}=\Omega_{\text{A}}^2,~~\tilde B^{\text{B}}=-2\Omega_{\text{A}}\kappa_{\text{A}},~~\tilde C^{\text{B}}=\kappa_{\text{A}}^2.
\label{parB}
\end{eqnarray}
Next expanding the sines in (\ref{d1}) we have
\begin{eqnarray}
&&D^{\text{A},\text{B}}=a^{\text{A},\text{B}}\sin Np-b^{\text{A},\text{B}}\cos Np,
\label{d2}
\\
&&a^{\text{A},\text{B}}=\tilde A^{\text{A},\text{B}}\cos p+\tilde B^{\text{A},\text{B}}\cos 2p+\tilde C^{\text{A},\text{B}}\cos 3p,
\label{aa}
\\
&&b^{\text{A},\text{B}}=\tilde A^{\text{A},\text{B}}\sin p+\tilde B^{\text{A},\text{B}}\sin 2p+\tilde C^{\text{A},\text{B}}\sin 3p.
\label{bb}
\end{eqnarray}
Further, expanding the norm squared, we finally obtain
\begin{eqnarray}
|D^{\text{A},\text{B}}|^2=\frac{1}{2}\bigg[L^{\text{A},\text{B}}-M^{\text{A},\text{B}}\cos 2Np-C^{\text{A},\text{B}}\sin 2Np\bigg],
\label{d3}
\end{eqnarray}
or
\begin{eqnarray}
|D^{\text{A},\text{B}}|^2=\frac{1}{2}\bigg[L^{\text{A},\text{B}}-K^{\text{A},\text{B}}\cos (2Np-\phi^{\text{A},\text{B}})\bigg],
\label{d4}
\end{eqnarray}
where we have introduced the parameters
\begin{eqnarray}
&&L^{\text{A},\text{B}}=|a^{\text{A},\text{B}}|^2+|b^{\text{A},\text{B}}|^2,
\label{L}
\\
&&M^{\text{A},\text{B}}=|a^{\text{A},\text{B}}|^2-|b^{\text{A},\text{B}}|^2,
\label{M}
\\
&&C^{\text{A},\text{B}}=a^{\text{A},\text{B}}(b^{\text{A},\text{B}})^\ast+(a^{\text{A},\text{B}})^\ast b^{\text{A},\text{B}},
\label{C} 
\\
&&K^{\text{A},\text{B}}=\sqrt{(M^{\text{A},\text{B}})^2+(C^{\text{A},\text{B}})^2},
\label{K}
\\
&&\tan\phi^{\text{A},\text{B}} = \frac{C^{\text{A},\text{B}}}{M^{\text{A},\text{B}}}.
\label{phi}
\end{eqnarray}
Since the norm of the cosine is less than one, it follows from
(\ref{d4}) that the upper and lower bounds of $|D^{\text{A},\text{B}}|^2$ are given by 
$(L^{\text{A},\text{B}}+K^{\text{A},\text{B}})/2$ and $(L^{\text{A},\text{B}}-K^{\text{A},\text{B}})/2$, respectively. 
Hence, from (\ref{tom}) it follows that the upper and lower bounds of $T^{\text{A},\text{B}}(\omega)$ are given by
\begin{eqnarray}
&&T^{\text{A},\text{B}}_{\text{max}}(\omega)=2\frac{(2\Gamma\omega\kappa_{\text{A}}\sin p)^2}{L^{\text{A},\text{B}}-K^{\text{A},\text{B}}},
\label{t1}
\\
&&T^{\text{A},\text{B}}_{\text{envelope}}(\omega)=2\frac{(2\Gamma\omega\kappa_{\text{A}}\sin p)^2}{L^{\text{A},\text{B}}+K^{\text{A},\text{B}}},
\label{t2}
\end{eqnarray}
respectively.  
The final step in obtaining a large $N$ expression for the CGF is achieved by using the
integral \cite{Gradshteyn65}
\begin{eqnarray}
\int_0^{2\pi}\frac{dp}{2\pi}
\ln(a+b\cos p)=\ln\left[\frac{a+\sqrt{a^2-b^2}}{2}\right],
\label{int3}
\end{eqnarray}
and ignoring the phase shift $\phi$, which  yields a subdominant contribution in the large
$N$ limit. Integrating over the $N$ dependent oscillations and including the contributions
from each phonon sub band we obtain the following large $N$ expression for CGF:
\begin{eqnarray}
\tilde\mu^{\text{A},\text{B}}(\lambda)=-\int_0^{\pi}\frac{dp}{2\pi}\sum_{n=1,2}
\rho_n\ln\left[\frac{L^{\text{A},\text{B}}_n+B_n+
\sqrt{(L^{\text{A},\text{B}}_n+B_n)^2-(K^{\text{A},\text{B}}_n)^2}}{L^{\text{A},\text{B}}_n+\sqrt{(L^{\text{A},\text{B}}_n)^2-(K^{\text{A},\text{B}}_n)^2}}\right]
\label{cgf4}
\end{eqnarray}
where $B_n=2(2\Gamma\omega_n\kappa_{\text{A}}\sin p)^2 f(\lambda)$, $L_n=L(\omega_n)$, and
$K_n=K(\omega_n)$ for n=1,2.
This is our main result which we proceed to discuss in the next section.
\section{\label{dis} Discussion}
\subsection{Cumulant generating function for large $N$}
Here we turn to a discussion of the main result, namely the expression (\ref{cgf4})
for the CGF in the large $N$ limit. First we note that since $f(\lambda)=0$ for 
$\lambda=0$ the CGF locks onto zero, i.e., $\tilde\mu(0)=0$, as required by
normalisation. Moreover, detailed analysis shows that to leading order in
$\mu$, i.e., $(d\mu(\lambda)/d\lambda)_{\lambda=0}=\langle Q\rangle/t$, the integral
in (\ref{cgf4}) can be carried out analytically. The corresponding expressions
for the heat current $\langle Q\rangle/t$ are in agreement with the result 
obtained by Kannan et al. \cite{Kannan11}. In the general case we have been unable
to reduce the complex expression (\ref{cgf4}) further.

Based on the finite $N$ expression for the CGF in (\ref{cgf3}) we have in Fig.~\ref{fig4} 
plotted the contributions to the CGF arising from the acoustic
and optical phonons, respectively, for $N=10$, $T_1=1$, and $T_2=1$.  In this case
the CGF is symmetrical. We note that the CGF is a downward convex function passing
through the origin with branch points at $\lambda_\pm=\pm 1$. For $T_1\neq T_2$ the
CGF is shifted and will pass through the origin for $\lambda=0$ and $\lambda=1/T_1-1/T_2$,
consistent with the AFT in (\ref{aft}). In Fig.~\ref{fig5} we have superimposed a plot of
of the $N\rightarrow\infty$ expression for the CGF,  $\tilde\mu$, on a plot of $\mu$
for $N=10$ and $T_1=T_2=1$.  We obtain an excellent fit indicating that the asymptotic 
large $N$ regime is attained already for small values of $N$.
\subsection{Transmission matrix}
The transmission matrix $T^{\text{A},\text{B}}$ given by (\ref{tom}) is an essential 
ingredient in the evaluation of the CGF. Owing to the resonance structure
in $G_{1N}^{\text{A,B}}$ the transmission matrix exhibits an oscillatory structure with period of order
$1/N$. The maximum value and lower envelope is given by (\ref{t1}) and (\ref{t2}), respectively.
In the large $N$ limit the oscillations merge together and allows for the smooth large $N$ approxiation
implemented in Sec.~\ref{cum}. Further analysis shows that in case B (odd), 
where the $u$ sublattice is driven by the heat reservoirs and the motion of $w$ sublattice is slaved to the
motion of the $u$ sublattice, see Fig.~\ref{fig1}, $T^{\text{B}}_{\text{max}}(\omega)=1$
for all $\omega$.  In case A (even) the transmission matrix $T^{\text{A}}(\omega)$ exhibits a 
similar form except for a shift of the oscillatory pattern owing to the altered boundary conditions.
Here the heat reservoirs drive each sublattice and only for the acoustic part do we have 
$T^{\text{A}}_{\text{max}}(\omega)=1$. 

In Fig.~\ref{fig6} we have depicted $T^{\text{B}}(\omega)$
as a function of $\omega$ for $N=5$, $m=1$, $M=2$, and $\kappa=1$.
The plot clearly shows the gap between the acoustic and optical phonons, the oscillatory structure 
being due to the resonance structure in $D^{\text{B}}(\omega)$. We have also plotted the lower envelope.
Finally we notice that $T^{\text{B}}(\omega)$ is bounded from above by unity. 
\subsection{Large deviation function}
Here we briefly discuss some of the implication for the large deviation function. Inserting the expression for the characteristic heat function (\ref{charheat}) in (\ref{pdis}) we obtain at long times
the following expression for the heat distribution.
\begin{eqnarray}
P(Q,t)\sim\int_{-i\infty}^{i\infty}\frac{d\lambda}{2\pi i}
\exp(-\lambda Q)\exp(t\mu(\lambda)).           
\label{pdis2}
\end{eqnarray}
This expression can be analyzed either as a Laplace transform or by a numerical simulation,
see \cite{Fogedby12}.
We shall not pursue such an approach here but note that a standard steepest descent 
argument or a Legendre transform \cite{Touchette09} implies that $P(Q,t)$ has the long time scaling form
\begin{eqnarray}
P(Q,t)\sim\exp(tF(Q/t)), 
\label{pdis3}
\end{eqnarray}
where the large deviation function $F(Q/t)$ (LDF) is determined by
\begin{eqnarray}
F(Q/t)=\mu(\lambda^\ast)-\lambda^\ast\mu'(\lambda^\ast).
\label{pldf}
\end{eqnarray}
Here  $\lambda^\ast$ is determined by the saddle point condition
\begin{eqnarray}
\mu'(\lambda^\ast)=Q/t.
\label{spc}
\end{eqnarray}
For the LDF the AFT for $\mu(\lambda)$ in (\ref{aft}) implies 
\begin{eqnarray}
F(Q/t)-F(-Q/t)=-(Q/t)(\beta_1-\beta_2). 
\label{ft2}
\end{eqnarray}
By inspection of the general expression (\ref{cgf2}) for the CGF
we infer that $\mu(\lambda)$ has the form of a downward
convex function passing through the origin $\mu(0)=0$ due to
normalization and through $\mu(\beta_1-\beta_2)=0$ owing to the
fluctuation theorem. The branch points $\lambda_\pm$ are determined by
the condition $1+T^{\text{A,B}}(\omega)f(\lambda)=0$, i.e., , the point where the log
diverges. Since  $T^{\text{A,B}}(\omega)\leq 1$, a little analysis shows
that the branch point are given by
\begin{eqnarray}
&&\lambda_+=\beta_1, 
\label{brp1}
\\
&&\lambda_-=-\beta_2.
\label{brp2}
\end{eqnarray}
Deforming the contour in the integral (\ref{pdis2}) to pass along
the real axis we pick up branch cut contributions in
$\mu(\lambda)$. Heuristically, we conclude that for large $|Q/t|$ 
the LDF depends linearly on $Q/t$, i.e.,
\begin{eqnarray}
&&F(Q/t)\sim -\lambda_+Q/t,~~~~\text{for}~Q/t\gg 0~,
\label{Fp}
\\
&&F(Q/t)\sim -|\lambda_-||Q/t|,~\text{for}~Q/t\ll 0 
\label{Fn};
\end{eqnarray}
where $\lambda_+$ and $\lambda_-$ have been  defined above. The
heat distribution thus exhibits exponential tails for large $|Q/t|$,
i.e.,
\begin{eqnarray}
&&P(Q/t)\propto\exp(-\lambda_+Q)~\text{for}~Q/t\gg 0, \label{ppp}
\\
&&P(Q/t))\propto\exp(-|\lambda_-||Q|)~\text{for}~Q/t\ll 0, \label{pnn}
\end{eqnarray}
with $\lambda_+$ and $\lambda_-$ given by (\ref{brp1}) and
(\ref{brp2}). It is interesting that the tails in the $Q$ distribution are determined only
by the reservoir temperatures. 
\section{\label{con}Conclusion}
In the present paper we have extended our previous work on the cumulant 
generating function and the large deviation function for the simple harmonic chain
to the case of an alternating mass chain. From a technical point of
view the analysis is more complex due to the two band structure arising
from the acoustic and optical phonon branches. We find that the transmission
matrix exhibits a two band structure. These results are in complete agreement with
Kannan el al \cite{Kannan11}. The contributions from the two
branches to the cumulant generating function are also identified.
Finally, we have extended the large $N$ approximation in \cite{Fogedby12}. 
We find that the cumulant generating function and thus the  heat
and higher cumulants of the heat are manifestly independent of the system size 
$N$ for large $N$. We find that the independence of $N$ sets in already at small $N$.
This is consistent with the fact that the system does not attain local equilibrium
and that Fourier's law does not hold. Finally, we have confirmed that the results
depend on whether the chain is composed of an even (case A) or odd (case B) number of
particles.
\acknowledgements 
We are grateful to A. Imparato for interesting discussions.
This work has been supported by a grant from The Danish Research Council.
\appendix
\section*{Appendix: Green's functions}
\begin{figure}
\includegraphics[width=0.7\hsize]{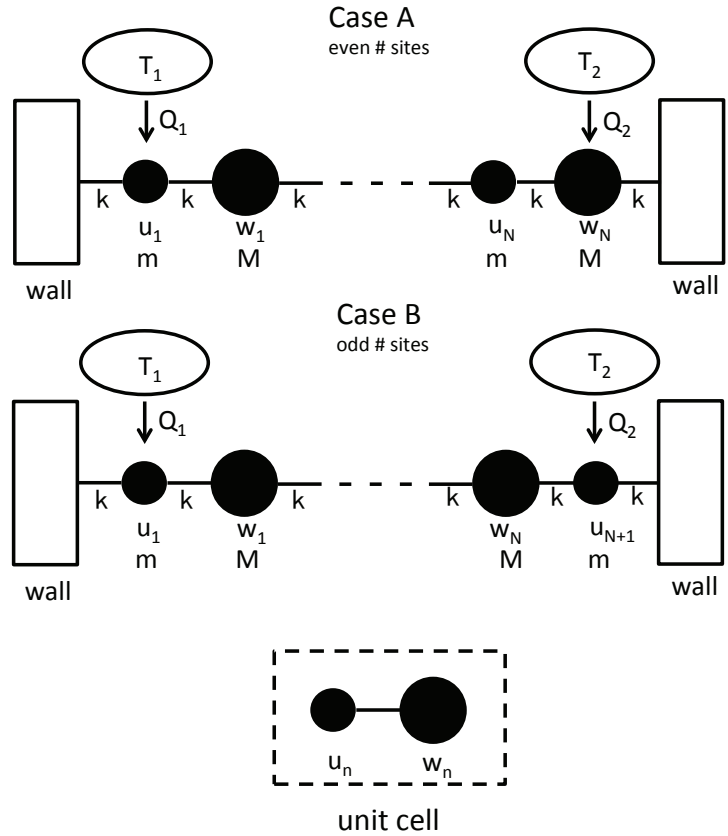}
\caption{We depict the two possible configuration for the
alternating mass chain and the appropriate
unit cell with basis. In case A we have an integer
set of unit cells, each containing a mass $m$ particle
with displacement $u_n$ and a mass $M$ particle
with displacement $w_n$. The particles are attached
by springs with spring constant $\kappa$. Particle $u_1$
is driven by a reservoir at temperature $T_1$; particle
$w_n$ driven at $T_2$. In case B particle $u_{N+1}$ in
a half filled unit cell is driven at $T_2$.} \label{fig1}
\end{figure}

\begin{figure}
\includegraphics[width=1.0\hsize]{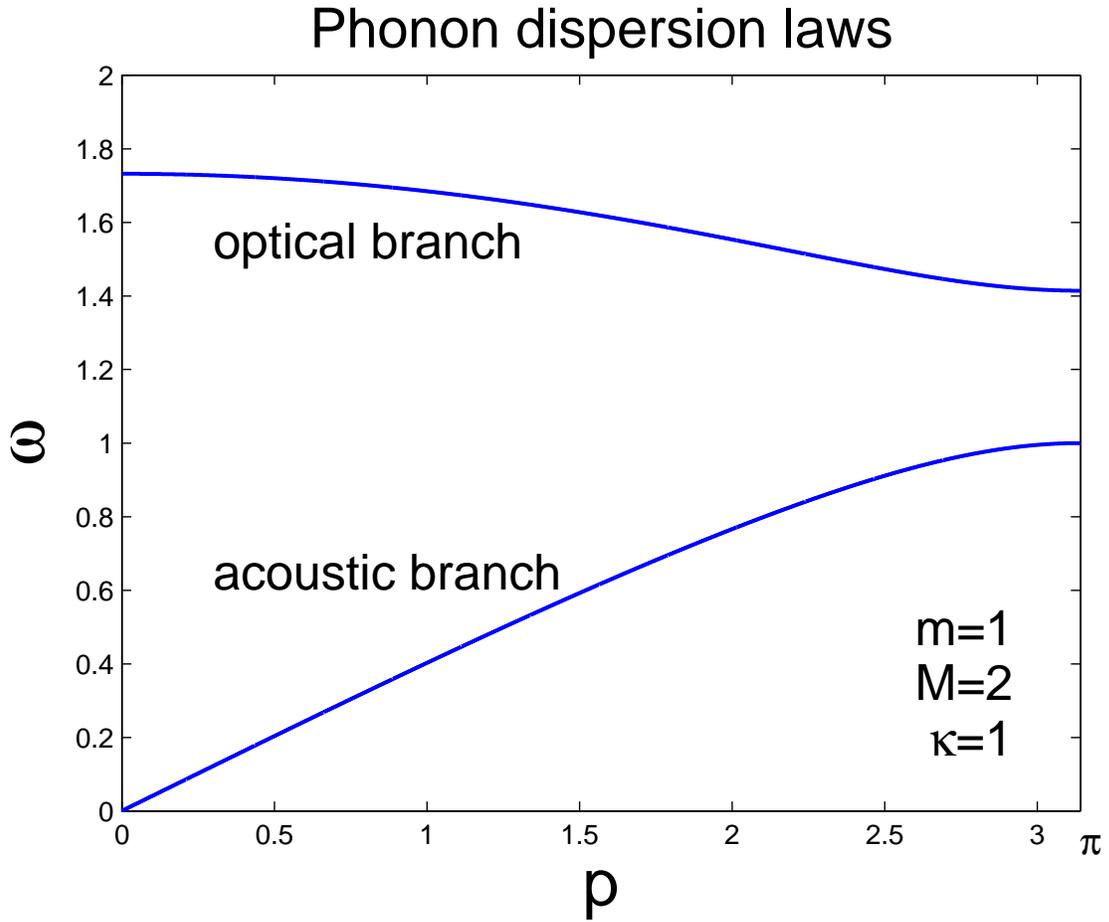}
\caption{We depict the acoustic and optical phonon
branches in a plot of $\omega$ versus $p$ given in
(\ref{ac}), (\ref{op}), and (\ref{ss}). The
wavenumber range is $0<p<\pi$. We have set $m=1$,
$M=2$, and $\kappa=1$.} \label{fig2}
\end{figure}

\begin{figure}
\includegraphics[width=1.0\hsize]{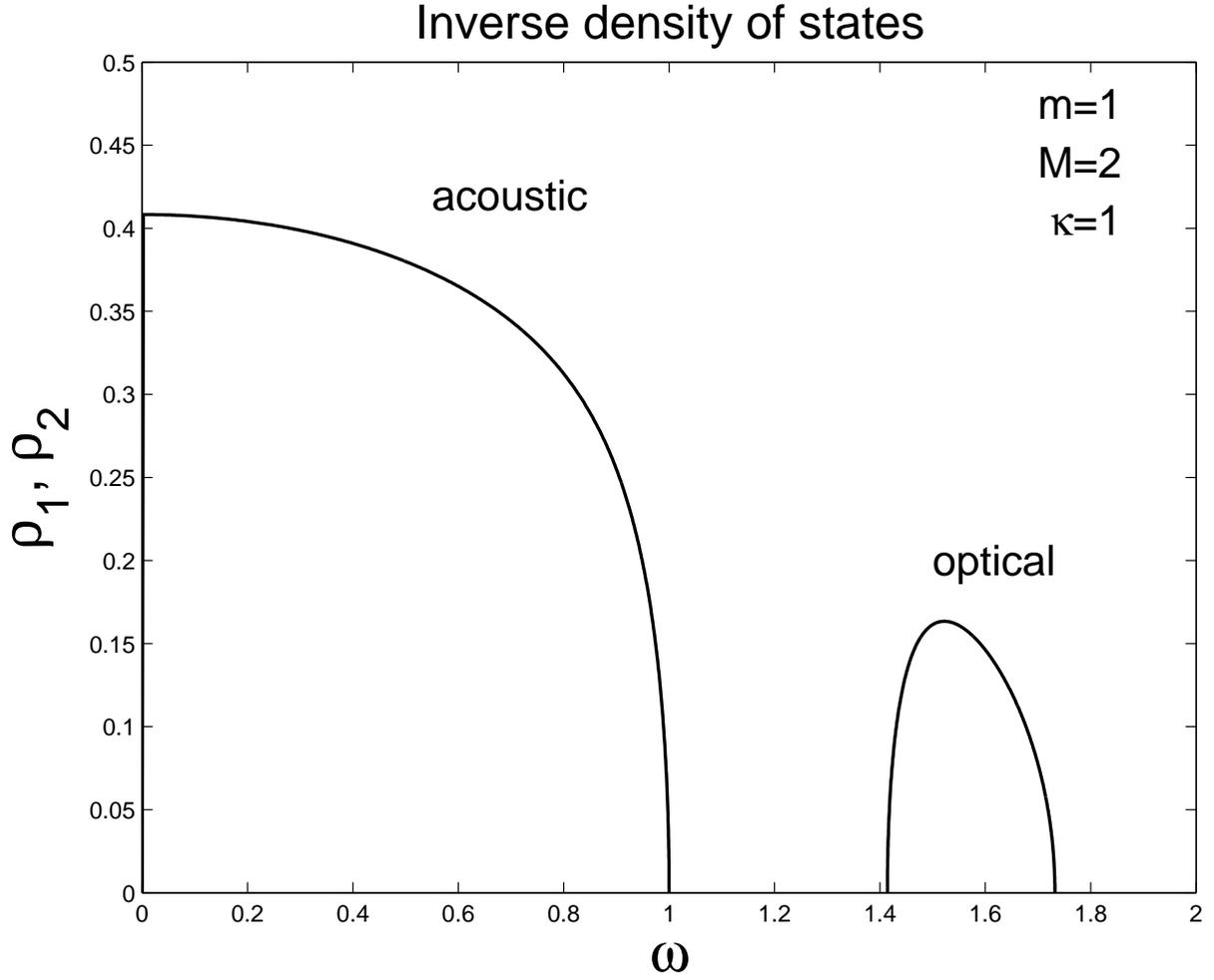}
\caption{We depict the inverse density of
states, $\rho_n=d\omega_n/dp$, for
the acoustic and optical branches,
respectively, given by (\ref{invdens1}) and (\ref{invdens2}). We plot $\rho_n$ as function
of $\omega$ in order to exhibit the band gap.
We have set $m=1$,
$M=2$, and $\kappa=1$.} 
\label{fig3}
\end{figure}

\begin{figure}
\includegraphics[width=1.0\hsize]{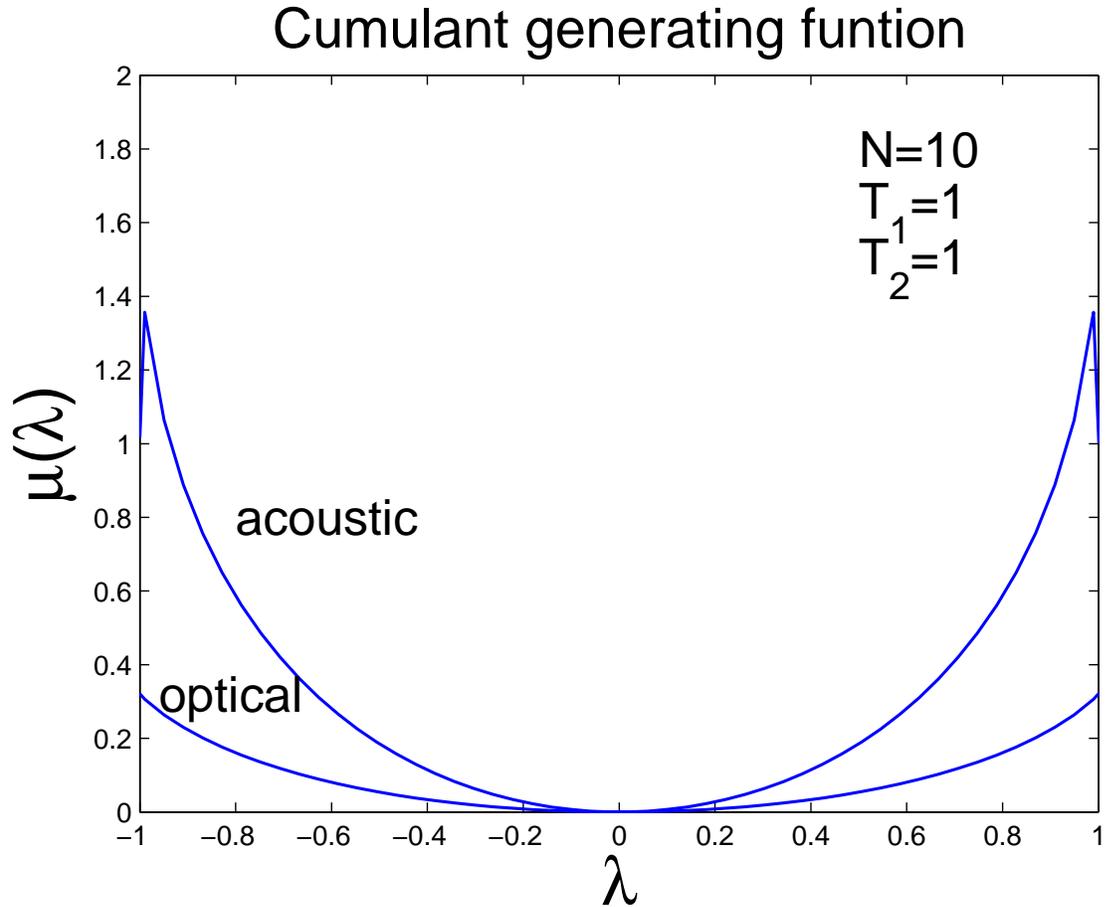}
\caption{We depict the contributions to the cumulant generating function
from the acoustic phonons, upper branch, and the optical phonons,
the lower branch, as function of $\lambda$, based on the expression
(\ref{cgf3}). We set $N=10$ and $T_1=T_2=1$, yielding
a symmetrical CGF. The branch points are at $\lambda_\pm=\pm 1$.} \label{fig4}
\end{figure}

\begin{figure}
\includegraphics[width=1.0\hsize]{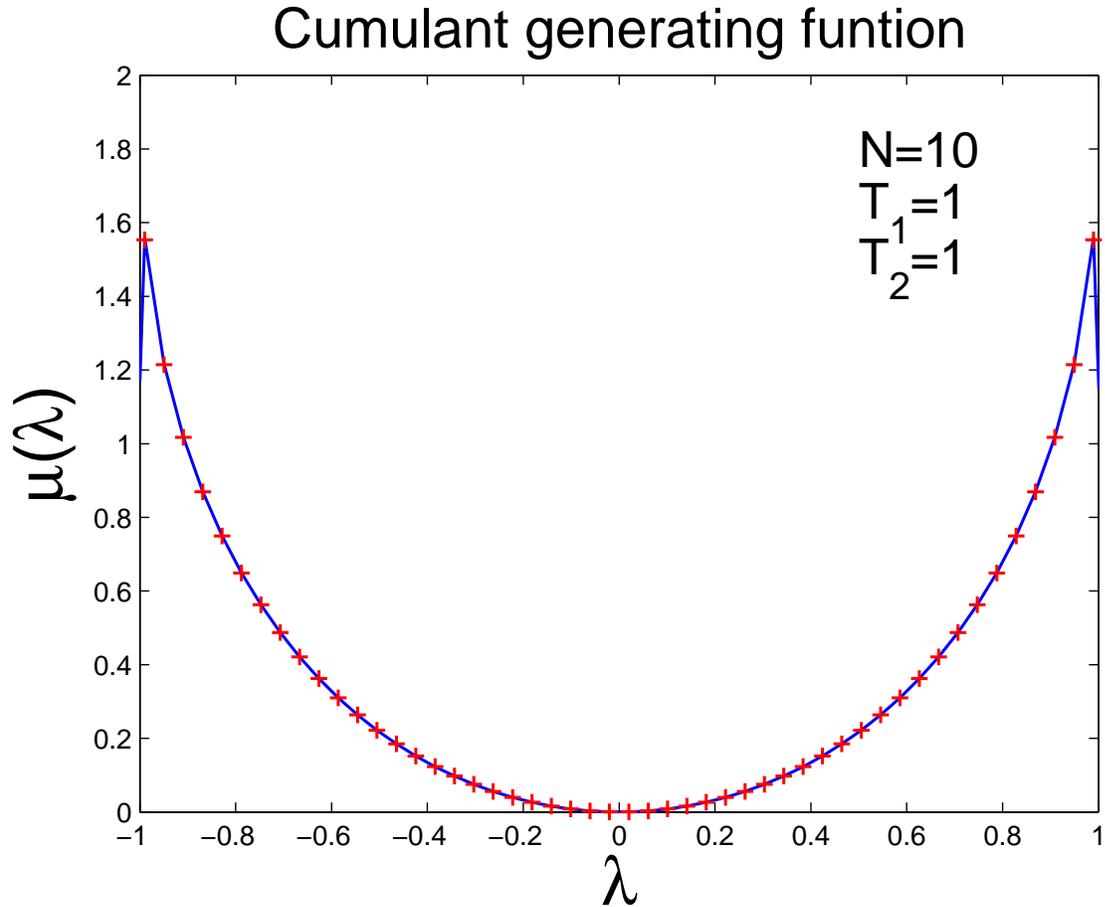}
\caption{We depict the full cumulant generating function, $\mu$, including
both the acoustic and optical contributions, given by (\ref{cgf3}) . Superimposed,
indicated by red crosses, we have plotted the large $N$ approximation of CGF, $\tilde\mu$,
given by (\ref{cgf4}). We have set $N=10$, $T_1=1$, and $T_2=1$. We
find excellent agreement.} \label{fig5}
\end{figure}

\begin{figure}
\includegraphics[width=1.0\hsize]{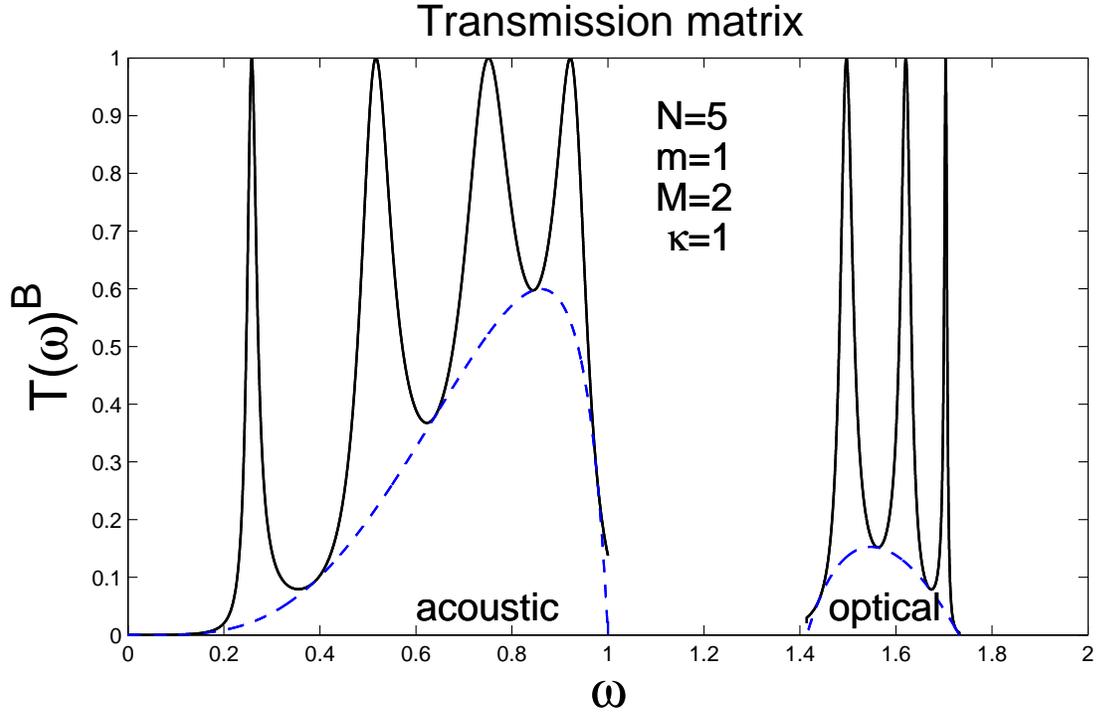}
\caption{We depict the transmission matrix
$T^{\text{B}}(\omega)$, given by (\ref{tom}), as function of $\omega$
for $N=5$, $m=1$, $M=2$, and $\kappa=1$ in the case B (odd). The oscillatory structure arises
from the resonance structure in $D^{\text{B}}(\omega)$.
The dashed line envelope arises from a large
$N$ approximation derived in appendix. The upper bound
is $T^{\text{B}}(\omega)=1$.} \label{fig6}
\end{figure}

A basic ingredient in our analysis are the Green's functions $G$ and $F$ in (\ref{un}) and (\ref{wn})
describing the propagation of lattice vibrations across the chain. In Kannan et al. \cite{Kannan11} the
derivation of the Green's functions is done using a determinantal approach, here we derive them directly
from the equations of motion (\ref{eq11}) and (\ref{eq22}) together with (\ref{c1a}) to (\ref{c2b}). 
The scheme follows the method used in \cite{Fogedby12} with the added complications due to
the two band structure. We obtain for the $u$ and $w$ sub lattices in the asymmetrical  case A, see Fig.~\ref{fig1},
the equations of motion
\begin{eqnarray}
&&\Omega_{\text{A}} u_1=\kappa_{\text{A}} u_2+\xi_1,
\label{a1}
\\
&&\Omega_{\text{C}} u_N=\kappa_{\text{D}} u_{N-1}+\xi_N,
\label{a2}
\\
&&\Omega_{\text{D}} w_1=\kappa_{\text{C}} w_2+\xi_1,
\label{a3}
\\
&&\Omega_{\text{B}} w_N=\kappa_{\text{B}} w_{N-1}+\xi_N.
\label{a4}
\end{eqnarray}
Likewise, in the symmetrical case B, see see Fig.~\ref{fig1},
the equations of motion
\begin{eqnarray}
&&\Omega_{\text{A}} u_1=\kappa_{\text{A}} u_2+\xi_1,
\label{b1}
\\
&&\Omega_{\text{A}} u_{N+1}=\kappa_{\text{A}} u_N+\xi_N,
\label{b2}
\\
&&\Omega_{\text{D}} w_1=\kappa_{\text{C}} w_2+\xi_1,
\label{b3}
\\
&&\Omega_{\text{D}} w_N=\kappa_{\text{C}} w_{N-1}+\xi_N.
\label{b4}
\end{eqnarray}
We have introduced the parameters 
\begin{eqnarray}
&&\Omega_{\text{A}}=\Omega_1-\kappa^2/\tilde\Omega_2,
\label{p1}
\\
&&\Omega_{\text{B}}=\Omega_2-\kappa^2/\tilde\Omega_1,
\label{p2}
\\
&&\Omega_{\text{C}}=(\Omega_2/\kappa)(\tilde\Omega_1-\kappa^2/\tilde\Omega_2)-\kappa,
\label{p3}
\\
&&\Omega_{\text{D}}=(\Omega_1/\kappa)(\tilde\Omega_2-\kappa^2/\tilde\Omega_1)-\kappa,
\label{p4}
\\
&&\kappa_{\text{A}}=\kappa^2/\tilde\Omega_2,
\label{p5}
\\
&&\kappa_{\text{B}}=\kappa^2/\tilde\Omega_1,
\label{p6}
\\
&&\kappa_{\text{C}}=\kappa\Omega_1/\tilde\Omega_1,
\label{p7}
\\
&&\kappa_{\text{D}}=\kappa\Omega_2/\tilde\Omega_2,
\label{p8}
\end{eqnarray}
where $\tilde\Omega_1$, $\tilde\Omega_2$,  $\Omega_1$, and $\Omega_2$ are 
given by (\ref{tom1}), (\ref{tom2}), (\ref{om1}), and (\ref{om2}).
Note that unlike the simple harmonic chain the parameters here acquire
an explicit $\omega$ dependence due to the dynamical coupling of
the two sub lattices.

Searching for plane wave solutions of the form
\begin{eqnarray}
&&u_n=\alpha_1\exp(ipn)+\beta_1\exp(-ipn),
\label{a}
\\
&&w_n=\alpha_2\exp(ipn)+\beta_2\exp(-ipn),
\label{b}
\end{eqnarray}
the coefficients $\alpha$ and $\beta$ are readily determined 
by insertion in the equations of motion.

\noindent
Case A:
\begin{eqnarray}
&&G_{n1}^{\text{A}}=\frac{\Omega_{\text{C}}\sin(N-n)p-\kappa_{\text{D}}\sin(N-1-n)p}{D_1^{\text{A}}},
\label{g1a}
\\
&&G_{nN}^{\text{A}}=\frac{\Omega_{\text{A}}\sin(n-1)p-\kappa_{\text{A}}\sin(n-2)p}{D_1^{\text{A}}},
\label{g2a}
\\
&&F_{n1}^{\text{A}}=\frac{\Omega_{\text{B}}\sin(N-n)p-\kappa_{\text{B}}\sin(N-1-n)p}{D_2^{\text{A}}},
\label{f1a}
\\
&&F_{nN}^{\text{A}}=\frac{\Omega_{\text{D}}\sin(n-1)p-\kappa_{\text{C}}\sin(n-2)p}{D_2^{\text{A}}},
\label{fa2}
\\
&&D_1^{\text{A}}=\Omega_{\text{A}}\Omega_{\text{C}}\sin(N-1)p-(\Omega_{\text{A}}\kappa_{\text{D}}+\Omega_{\text{C}}\kappa_{\text{A}})\sin(N-2)p+\kappa_{\text{A}}\kappa_{\text{D}}\sin(N-3)p,
\label{d1a}
\\
&&D_2^{\text{A}}=\Omega_{\text{B}}\Omega_{\text{D}}\sin(N-1)p-(\Omega_{\text{D}}\kappa_{\text{B}}+\Omega_{\text{B}}\kappa_{\text{C}})\sin(N-2)p+\kappa_{\text{B}}\kappa_{\text{C}}\sin(N-3)p.
\label{d2a}
\end{eqnarray}
Case B:
\begin{eqnarray}
&&G_{n1}^{\text{B}}=\frac{\Omega_{\text{A}}\sin(N-n)p-\kappa_{\text{A}}\sin(N-1-n)p}{D_1^{\text{B}}},
\label{g1b}
\\
&&G_{nN}^{\text{B}}=\frac{\Omega_{\text{A}}\sin(n-1)p-\kappa_{\text{A}}\sin(n-2)p}{D_1^{\text{B}}},
\label{g2b}
\\
&&F_{n1}^{\text{B}}=\frac{\Omega_{\text{D}}\sin(N-n)p-\kappa_{\text{C}}\sin(N-1-n)p}{D_2^{\text{B}}},
\label{f1b}
\\
&&F_{nN}^{\text{B}}=\frac{\Omega_{\text{D}}\sin(n-1)p-\kappa_{\text{C}}\sin(n-2)p}{D_2^{\text{B}}},
\label{f2b}
\\
&&D_1^{\text{B}}=\Omega_{\text{A}}^2\sin(N-1)p-2\Omega_{\text{A}}\kappa_{\text{A}}\sin(N-2)p+\kappa_{\text{A}}^2\sin(N-3)p,
\label{d1b}
\\
&&D_2^{\text{B}}=\Omega_{\text{D}}^2\sin(N-1)p-2\Omega_{\text{D}}\kappa_{\text{C}}\sin(N-2)p+\kappa_{\text{C}}^2\sin(N-3)p.
\label{d2b}
\end{eqnarray}
We have in particular
\begin{eqnarray}
&&G_{1N}^{\text{A}}=\frac{\kappa_{\text{A}}\sin p}{D_1^{\text{A}}},
\label{g1Na}
\\
&&G_{1N}^{\text{B}}=\frac{\kappa_{\text{A}}\sin p}{D_1^{\text{B}}},
\label{g1Nb}
\end{eqnarray}
Detailed inspection of the above results for $G$ and $F$ shows that they are in agreement with the determinantal
results in \cite{Kannan11}.

\newpage
\end{document}